\documentclass[conference, 12pt]{IEEEtran}  

\IEEEoverridecommandlockouts
\usepackage{cite}
\usepackage{amsmath,amssymb,amsfonts}
\usepackage{algorithmic}
\usepackage{graphicx}
\usepackage{textcomp}
\usepackage{xcolor}
\usepackage{multirow}

\def\BibTeX{{\rm B\kern-.05em{\sc i\kern-.025em b}\kern-.08em
    T\kern-.1667em\lower.7ex\hbox{E}\kern-.125emX}}
\begin{document}

\title{Efficient Software Vulnerability Detection Using Transformer-based Models\\
}

\author{
\IEEEauthorblockN{Sameer Shaik}
\IEEEauthorblockA{\textit{DePaul University} \\
Chicago, Illinois, USA \\
sameershaik65@gmail.com}

\and

\IEEEauthorblockN{Zhen Huang}
\IEEEauthorblockA{\textit{DePaul University} \\
Chicago, Illinois, USA \\
zhen.huang@depaul.edu}

\and
\IEEEauthorblockN{Daniela Stan Raicu}
\IEEEauthorblockA{\textit{DePaul University} \\
Chicago, Illinois, USA \\
draicu@cdm.depaul.edu}

\and
\IEEEauthorblockN{Jacob Furst}
\IEEEauthorblockA{\textit{DePaul University} \\
Chicago, Illinois, USA \\
jfurst@cdm.depaul.edu}

}

\maketitle

\begin{abstract}

Detecting software vulnerabilities is critical to ensuring the security and reliability of modern computer systems. Deep neural networks have shown promising results on vulnerability detection, but they lack the capability to capture global contextual information on vulnerable code. To address this limitation, we explore the application of transformers for C/C++ vulnerability detection. We use program slices that encapsulate key syntactic and semantic features of program code, such as API function calls, array usage, pointer manipulations, and arithmetic expressions. By leveraging transformers' capability to capture both local and global contextual information on vulnerable code, our work can identify vulnerabilities accurately. Combined with data balancing and hyperparameter fine-tuning, our work offers a robust and efficient approach to identifying vulnerable code with moderate resource usage and training time. 
\end{abstract}


\section{Introduction}
Software vulnerabilities are a critical and ongoing challenge to the security and stability of software systems. Despite significant advancements in secure programming practices, vulnerabilities such as memory mismanagement, pointer misuse, and arithmetic errors continue to pose severe risks. These weaknesses, whether in application logic or system configurations, are often exploited by adversaries to compromise system integrity, gain unauthorized access, or disrupt operations, thus threatening both network and information security.

Over the years, many non-AI-based techniques have been employed for detecting vulnerabilities. Targeted symbolic execution is particularly effective in identifying use-after-free (UAF) vulnerabilities by analyzing the sequence of memory allocation (malloc), deallocation (free), and usage operations. Similarly, static taint analysis has been used to detect integer overflow vulnerabilities by tracing arithmetic operations at both def-site and use-site locations. Symbolic execution can leverage safety properties to flag violations in software security issues caused by vulnerabilities. However, these traditional methods, while useful in specific scenarios, often lack scalability, require significant manual effort, and may not generalize well to more complex or diverse vulnerability types.

In recent years, the emergence of deep learning techniques has opened new avenues for automatic vulnerability detection. VulDeePecker, for instance, utilizes BiLSTM (Bidirectional Long Short-Term Memory) network trained on slices of program code, which capture key segments of vulnerable code. Prior work also explored BiGRU (Bidirectional Gated Recurrent Unit) models that showed improved performance over BiLSTM. Although contextual information used by neural networks like BiLSTM and BiGRU, trained on program slices, can improve vulnearbility detection accuracy, they lack a mechanism to preserve longer contextual relationships in the code. Accurately detecting vulnerabilities requires models capable of understanding both local and long-range dependencies, with which BiLSTM struggles due to its limited context window.

Recognizing these limitations, recent research has shifted toward transformer-based architectures because transformer-based models, such as BERT and its variants, have shown great potential to address this challenge. These models excel at capturing both local and long-range dependencies, making them particularly suitable for tasks such as vulnerability detection, where contextual information is crucial. LineVul uses a transformer-based model specifically designed to detect C/C++ function-level vulnerabilities. LineVul demonstrates a significant improvement in vulnerability detection accuracy due to its ability to retain contextual information across longer sequences of code. DistilBERT, a lightweight yet effective alternative to BERT, has also shown promising results for large-scale code analysis.

In this work, we focus on fine-tuning DistilBERT, CodeBERT, Google Gemma-2b  to classify vulnerable and non-vulnerable code. We train these models on program slices representing vulnerability-related constructs, specifically API calls, AU, PU, and AE, and employ data balancing techniques to address the inherent imbalance in vulnerability datasets, where non-vulnerable code typically outnumbers vulnerable code. By leveraging the strengths of transformer-based architectures, our work aims to overcome the limitations of previous approaches and improve the robustness and accuracy of vulnerability detection.

This paper makes the following major contributions:
\begin{itemize}
\item We explore two different approaches to balancing the vulnerable samples and non-vulnerable samples in our data set.
\item We fine-tune three different transformer models to achieve optimal vulnerability detection accuracy.
\item We measure the computational cost of training the three transformer models.
\end{itemize}

The remainder of this paper is organized as follows. Section~\ref{related} discusses related work, including traditional and modern approaches to vulnerability detection. Section~\ref{methodology} details our methodology, focusing on data preparation, model fine-tuning, and balancing techniques, and our implementation. Section~\ref{evaluation} presents our experimental setup and results. Section~\ref{discussion} discusses our results. Finally, we conclude in Section~\ref{conclusion}.

\section{Related Work}\label{related}

Over the years, several techniques leveraging machine learning and deep learning models have been proposed for detecting vulnerabilities in C/C++ programs. These approaches vary in terms of feature extraction methods, model architectures, and detection capabilities. In this section, we explore relevant studies, focusing on neural networks, transformer models, and program slicing techniques for vulnerability detection.

\subsection{Vulnerability Detection using Neural Networks}
Machine learning and deep learning models have gained prominence in vulnerability detection, often surpassing traditional methods such as pattern matching and code similarity detection. A notable example is the use of Bidirectional Gated Recurrent Units (BGRU), where a balance of vulnerable and non-vulnerable program slices was used to improve detection accuracy. In one such study, BGRU, paired with the ADAM optimizer, demonstrated superior performance in detecting vulnerabilities within C/C++ code, outperforming traditional techniques like pattern matching. Embedding models such as Word2Vec were used to transform program slices into vectors, revealing the diverse distribution of vulnerability-related constructs like API calls, array usage, and pointer manipulation. Dataset balancing was critical to achieving more accurate results, highlighting the need for strategies that handle class imbalance in vulnerability datasets effectively.

\subsection {VulDeePecker} 
VulDeePecker ~\cite{li2018vuldeepecker} introduced the concept of "code gadgets," which represent semantically related lines of code, for vulnerability detection using a BiLSTM model. This approach transformed source code into vector representations to detect vulnerabilities. However, the system encountered significant challenges, particularly high false negative rates, which limited its detection capabilities. VulDeePecker's dataset was derived from the National Vulnerability Database (NVD) and Software Assurance Reference Dataset (SARD). The process involved two phases: the extraction of code gadgets and API function calls, followed by their transformation into vectors for BiLSTM-based analysis. Despite its innovation, VulDeePecker struggled with false negatives, prompting the need for more robust methods.

\subsection {LineVul}
LineVul ~\cite{fu2022linevul} represents a major leap forward in vulnerability detection, employing a transformer-based model to address the limitations of earlier methods. Unlike project-specific approaches, LineVul utilizes the CodeBERT pre-trained model and employs self-attention mechanisms to capture long-range dependencies in the source code. By focusing on line-level predictions, LineVul provides more granular insights into potential vulnerabilities. It achieved impressive results, with an F-measure of 0.91 and a high precision rate of 97\%. This demonstrates the effectiveness of transformer-based models in capturing complex relationships in code, offering a substantial improvement over traditional methods like IVDetect.

\subsection{ SySeVR}
SySeVR ~\cite{li2021sysevr} introduces a systematic approach for detecting vulnerabilities in C/C++ programs by leveraging both syntactic and semantic features. The framework distinguishes between Syntax-based Vulnerability Candidates (SyVCs) and Semantics-based Vulnerability Candidates (SeVCs), converting these into vector representations that can be processed by deep neural networks. SySeVR's BGRU-based model outperformed state-of-the-art tools, demonstrating its ability to detect a wide range of vulnerabilities, including previously unknown ones. The dataset used was sourced from NVD and SARD. Despite its success, the framework faces challenges, such as adapting to other programming languages and improving its labeling methods to better handle issues like code duplication.

\subsection{ Program Slicing and Transformer-Based Models for Vulnerability Detection}
Program slicing has become an increasingly popular technique for capturing relevant features from source code, particularly when combined with transformer models like CodeBERTa. In recent work, slicing-based approaches outperformed function-level analysis in vulnerability detection tasks, achieving better accuracy by preserving the context of vulnerable code fragments. CodeBERTa, a transformer model based on BERT, was trained on a large corpus of code and fine-tuned to detect vulnerabilities in C/C++ programs. The use of program slicing allowed for more fine-grained detection, improving performance compared to function-level models ~\cite{10356694}.

\subsection{ Self-Attention-Based Vulnerability Detection}
Recognizing the limitations of RNN and BiLSTM models in understanding complex natural language structures, researchers have explored transformer models for vulnerability detection. A base transformer model, trained on code slices rather than full code blocks, was configured with a 500-token embedding vector and optimized using cross-entropy loss and the ADAM optimizer. This self-attention-based approach achieved an F1 score of 90\%, highlighting the capability of transformers in capturing intricate dependencies and retaining semantic information in code slices ~\cite{wu2021self}.

\subsection{ Transformer-Based Approaches for Vulnerability Detection}
RoBERTa, an optimized version of BERT, has demonstrated its effectiveness in vulnerability detection tasks, outperforming traditional static and dynamic analysis techniques that often suffer from false positives. In one study, RoBERTa was trained on a real-world dataset, 'ReVeal,' and tested using a stratified K-fold method to mitigate the effects of imbalanced data. With hyperparameter tuning, including batch size and learning rate optimization, RoBERTa achieved significant improvements in test accuracy, demonstrating the robustness of transformer-based models in this domain \cite{thapa2022transformer}.

\subsection{ VulBERTa}
 A transformer-based model trained on real-world vulnerability data sourced from GitHub and Draper, presents a state-of-the-art solution for vulnerability detection. VulBERTa uses RoBERTa's architecture and is fine-tuned with a custom tokenization pipeline using Byte Pair Encoding (BPE). By integrating an MLP (Multi-Layer Perceptron) and CNN (Convolutional Neural Network) with the transformer model, VulBERTa achieved state-of-the-art performance on benchmark datasets like CodeXGLUE and D2A. VulBERTa's \cite{hanif2022vulberta} success in down-stream tasks demonstrates its scalability and adaptability for real-time vulnerability detection in diverse codebases.

\section{Methodology}\label{methodology}

In this section, we describe the methodological steps taken to implement and evaluate the vulnerability detection models, including data preparation, down-sampling techniques, and model architectures. The key focus of our work is the comparison of various down-sampling strategies and transformer models to improve vulnerability detection in C/C++ code slices.

\subsection{Down Sampling Techniques}

Handling class imbalance is a critical issue when training machine learning models on unbalanced datasets. In our case, 80\% of the data is classified as non-vulnerable, while only 20\% is vulnerable. Without addressing this imbalance, the model would likely be biased toward the majority class. To tackle this, we explore down-sampling methods based on two hypotheses, ensuring diverse representation from the four subclasses of vulnerabilities: API Function Call, Array Usage, Pointer Usage, and Arithmetic Expression.

\subsubsection{Hypothesis 1: Equal Distribution by Subclass}

\begin{figure}[h]
    \centering
    \includegraphics[width=8cm, height=6cm]{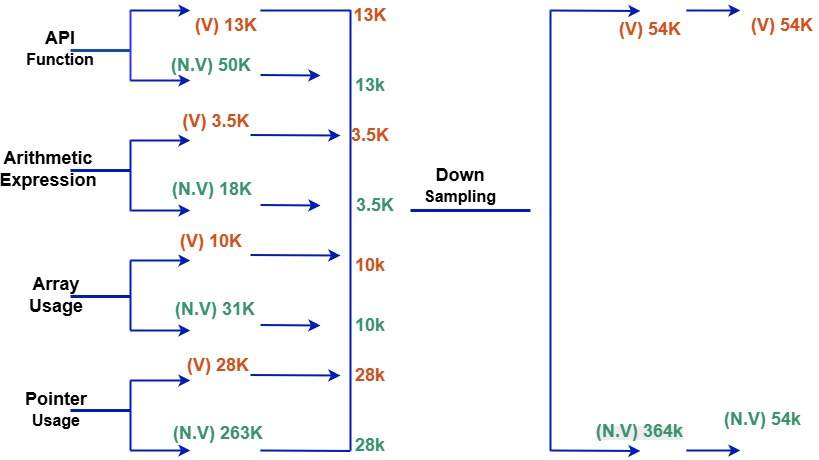}
    \caption{Down-sampling under Hypothesis 1: the numbers in orange refer to vulnerable samples, while the numbers in green refer to non-vulnerable samples.}
    \label{fig:hypothesis1}
\end{figure}

In Hypothesis 1, we addressed the class imbalance inherent in the dataset by constructing a balanced subset for each vulnerability category. The original dataset exhibited significant disparities between vulnerable and non-vulnerable code samples across four categories: API function Usage (API), array usage (AU), pointer usage (PU), and arithmetic expressions (AE). For instance, the pointer usage (PU) category contained 28,391 vulnerable samples but over 263,000 non-vulnerable samples, leading to potential bias in model training. To mitigate this, we selected all available vulnerable samples from each category and then randomly sampled an equal number of non-vulnerable samples within the same category. This ensured that each class was balanced independently, preserving vulnerability-specific features while preventing overrepresentation of safe code patterns. As a result, the final dataset under Hypothesis 1 comprised 56,395 vulnerable and 56,395 non-vulnerable samples, totaling 112,790 code instances. This balanced structure enabled a fair evaluation of each model’s ability to detect vulnerabilities without being skewed by class frequency. The down-sampling process under Hypothesis 1 is illustrated in Figure \ref{fig:hypothesis1}.

\subsubsection{Hypothesis 2: Balance by Smallest Subclass}

\begin{figure}[h]
    \centering
    \includegraphics[width=9cm, height=6cm]{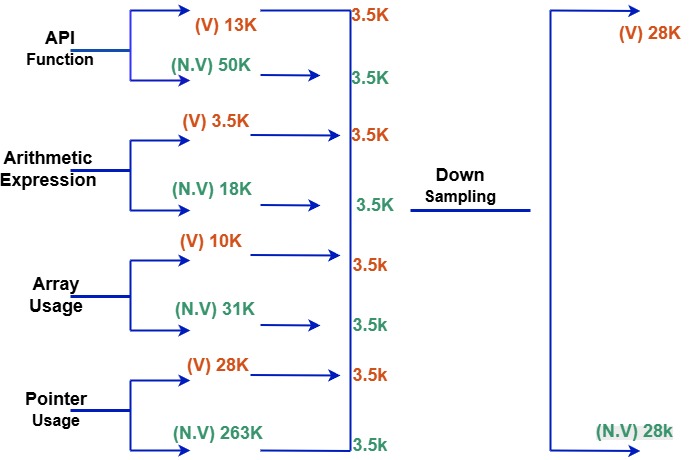}
    \caption{Down-sampling under Hypothesis 2: the numbers in orange refer to vulnerable samples, while the numbers in green refer to non-vulnerable samples.}
    \label{fig:hypothesis2}
\end{figure}

In Hypothesis 2, we adopted a more constrained downsampling approach to construct a highly uniform and minimal dataset for evaluating model performance under limited data conditions. Specifically, we selected the lowest available number of samples across all vulnerability categories, which was 3,475 vulnerable samples from the arithmetic expressions (AE) class. To maintain balance and prevent bias, we randomly selected an equal number of non-vulnerable samples, resulting in a final dataset of 6,950 code instances, equally split between vulnerable (3,475) and non-vulnerable (3,475) examples. These samples were evenly distributed across four categories—API Usage, array usage, pointer usage, and arithmetic expressions—ensuring that each vulnerability type was equally represented in both classes. This hypothesis is designed to test model effectiveness when data availability is minimal and perfectly balanced, providing insight into how well each model generalizes without relying on large or uneven training sets. The balanced data distribution is shown in Figure \ref{fig:hypothesis2}.

\subsection{Transformer Models for Vulnerability Detection}

Transformer models have proven to be highly effective for text-based tasks, such as classification, generation, and summarization, due to their ability to capture long-range dependencies in data through their attention mechanisms. In our work, we employ transformer models to classify C/C++ code slices as either vulnerable or non-vulnerable, leveraging their context-aware architecture. Transformers are particularly well-suited for this task as program code, although machine-readable, is structured similarly to natural language, where context plays a crucial role.

\subsubsection{DistilBERT: Lightweight Transformer by Hugging Face}
DistilBERT \cite{sanh2019distilbert} was developed by Hugging Face as a distilled version of the original BERT model by Google. It is based on transformer architecture, comprising 6 encoder layers (compared to BERT's 12), 768 hidden units, and 12 attention heads. DistilBERT uses knowledge distillation during training, where a smaller “student” model learns to approximate the outputs of a larger “teacher” model—in this case, BERT. This process enables the model to retain most of BERT’s performance while being significantly faster and less resource-intensive. In our vulnerability detection task, DistilBERT is fine-tuned on tokenized C/C++ program slices. Its attention mechanisms help capture complex dependencies such as unsafe pointer dereferencing or improper use of array indices across local and long-range contexts in code. This makes it a suitable candidate for real-time or edge-level static analysis tools.

\subsubsection{CodeBERT: Bimodal Code Understanding by Microsoft
}

CodeBERT \cite{feng2020codebert} was developed by Microsoft Research as a bimodal pre-trained transformer model for programming and natural languages. It is based on RoBERTa, a robustly optimized variant of BERT, and consists of 12 transformer encoder layers with 768 hidden dimensions and 12 attention heads. Unlike DistilBERT, CodeBERT is pre-trained on a corpus of code and associated documentation/comments from GitHub, spanning multiple programming languages including C, C++, Java, and Python. This bimodal training allows CodeBERT to learn both syntactic structure and semantic context in code. For our use case, CodeBERT is fine-tuned on labeled C/C++ code snippets that include vulnerability indicators such as unchecked buffer operations or unsafe memory manipulation. Its deep understanding of code-specific patterns makes it highly effective for capturing subtle vulnerabilities that are not easily detectable using traditional NLP models.

\subsubsection{GEMMA-2B: Efficient Decoder-Only Model by Google}

GEMMA-2B \cite{team2024gemma} is part of Google’s open-weight GEMMA (Generative Models from Google) family, released in 2024. Unlike BERT and CodeBERT, GEMMA-2B follows a decoder-only transformer architecture similar to GPT-style models, using causal self-attention layers. It consists of approximately 2 billion parameters, optimized for both instruction-following and efficient inference. Though GEMMA-2B is not specialized in code understanding, it has demonstrated versatility in various NLP tasks. We adapt GEMMA-2B for binary classification of vulnerable versus non-vulnerable code by fine-tuning it with instruction-style prompts and labeled C/C++ samples. Despite its generalist nature, GEMMA-2B's transformer blocks still enable it to model sequential token relationships, making it a useful model for assessing code safety in resource-constrained environments where smaller models are desirable.

\subsection{Training and Evaluation}
Once the datasets were down-sampled according to the two hypotheses, we split the data into training, validation, and test sets to evaluate the models. The training and validation sets were derived from the down-sampled data, while the test set was composed of the remaining samples. This ensured that the models were evaluated on a balanced test set, reflecting real-world conditions.


For Hypothesis 1, the data was split into 80\% training and 20\% validation using the \texttt{train\_test\_split} function from sklearn. For Hypothesis 2, due to the smaller dataset size after over-downsampling, we used a different strategy: test samples were drawn from the original dataset before down-sampling to ensure sufficient representation of all subclasses, except  for the vulnerable samples in the Arithmetic Expression subclass, which we used a 80\%-20\% split for training and validation.


\subsection{Model Architectures and Parameters}
We evaluate three transformer-based models for vulnerability detection: DistilBERT, CodeBERT, and Gemma-2B. Each model was fine-tuned using consistent hyperparameters to enable fair comparison across architectures.

\subsubsection{DistilBERT}
The DistilBERT model employs a distilled version of BERT architecture with 6 transformer layers instead of 12, maintaining 97\% of BERT's performance while being 40\% smaller. Key components:

\begin{itemize}
    \item \textbf{Architecture}: 6-layer transformer, 768 hidden dim, 12 attention heads
    \item \textbf{Parameters}: 66 million
    \item \textbf{Tokenizer}: WordPiece with 30,522 token vocabulary
    \item \textbf{Sequence Length}: 512 tokens (truncated/padded)
    \item \textbf{Fine-tuning}: Added classification head with 2 output neurons
\end{itemize}

We select it for its efficiency in processing code sequences while maintaining semantic understanding capabilities through self-attention mechanisms.

\subsubsection{CodeBERT}
Microsoft's CodeBERT  uses a 12-layer transformer pre-trained on both natural language and programming language (CodeSearchNet corpus):

\begin{itemize}
    \item \textbf{Architecture}: 12-layer transformer, 768 hidden dim, 12 attention heads
    \item \textbf{Parameters}: 125 million  
    \item \textbf{Tokenizer}: RoBERTa-based with 50,265 token vocabulary
    \item \textbf{Code-Specific Training}: MLM objective on bimodal data
    \item \textbf{Edge}: Captures code-specific patterns through code-nl alignment
\end{itemize}

We select it for its demonstrated effectiveness in code understanding tasks and its direct relevance to program analysis.

\subsubsection{Gemma-2B}
Google's Gemma-2B  employs a decoder-only transformer architecture optimized for code generation and understanding:

\begin{itemize}
    \item \textbf{Architecture}: 18-layer transformer, 2560 hidden dim, 32 attention heads
    \item \textbf{Parameters}: 2.5 billion
    \item \textbf{Tokenizer}: SentencePiece with 256k token vocabulary
    \item \textbf{Context Window}: 8192 tokens (limited to 512 for our task)
    \item \textbf{Specialization}: Instruction-tuned for code-related tasks
\end{itemize}

We select it as a state-of-the-art large language model to assess the scalability of vulnerability detection.

\subsubsection{Hyperparameters and Training}
Consistent training parameters across all models enable controlled comparison:

\begin{itemize}
    \item \textbf{Learning Rate}: $2 \times 10^{-5}$ (AdamW optimizer)
    \item \textbf{Batch Size}: 8 samples/device (gradient accumulation)
    \item \textbf{Training Epochs}: 5 (early stopping monitored)
    \item \textbf{Regularization}: Weight decay (0.01), dropout (0.1)
    \item \textbf{Tokenization}: Consistent padding/truncation to 512 tokens
\end{itemize}

The hyperparameters for fine-tuning DistilBERT, CodeBERT, and Gemma 2B were selected based on established practices for transformer-based models in code analysis tasks. A learning rate of \(2 \times 10^{-5}\) with the AdamW optimizer balances stable convergence and task-specific adaptation, as demonstrated in studies optimizing LLMs for vulnerability detection \cite{feng2020codebert,devlin2019bert}. A small batch size of 8 samples per device, coupled with gradient accumulation, mitigates memory constraints while preserving training stability \cite{chen2020training,gong2019efficient}. Training was limited to 5 epochs with early stopping to prevent overfitting, a strategy validated in prior work where transformer models converge efficiently on code datasets \cite{feng2020codebert,devlin2019bert}. Regularization via weight decay (0.01) and dropout (0.1) reduces model complexity, addressing the risk of overfitting in imbalanced vulnerability datasets \cite{liu2019roberta,lee2020regularization}. Code slices were tokenized and truncated to 512 tokens, ensuring critical syntactic and semantic features (e.g., pointer manipulations, API calls) are retained without exceeding typical transformer input limits \cite{guo2021graphcodebert,child2021long}. These choices collectively prioritize computational efficiency, generalization, and alignment with domain-specific best practices.s. All models were trained on NVIDIA A100 GPUs.

\subsection{Implementation Details}
Our implementation leverages the HuggingFace Transformers library \cite{wolf2019huggingface}:

\begin{itemize}
    \item \textbf{Data Preparation}: Program slices converted to HuggingFace Dataset format
    \item \textbf{Dynamic Padding}: DataCollatorWithPadding for memory efficiency
    \item \textbf{Metric Tracking}: Accuracy with evaluate library
    \item \textbf{Reproducibility}: Fixed random seeds (42), deterministic algorithms
\end{itemize}

The model architectures were particularly suited for capturing both local code patterns (through self-attention heads) and global semantic relationships (via transformer layers), crucial for identifying vulnerability signatures in code slices.

\subsubsection{Data Distribution}

Below is the count of data distributions for the \textbf{Hypothesis 1 and Hypothesis 2} down-sampling strategy:

\begin{table}[h]
\centering
\renewcommand{\arraystretch}{1.3}
\caption{Data Distribution for Hypothesis 1 and Hypothesis 2}
\label{table:data_distribution}
\begin{tabular}{|p{0.6cm}|r|r|r|r|r|}
\hline
\textbf{Kind} & \textbf{Data} & \textbf{VUL} & \textbf{NON VUL} & \textbf{Hypo\_1} & \textbf{Hypo\_2} \\
\hline
API & 15.30\% & 13,603 & 50,800 & 27,206 & 6,950 \\
\hline
AU  & 10\%    & 10,926 & 31,303 & 21,852 & 6,950 \\
\hline
PU  & 69.40\% & 28,391 & 263,450 & 56,782 & 6,950 \\
\hline
AE  & 5.30\%  & 3,475  & 18,679  & 6,950  & 6,950 \\
\hline
\textbf{Total} &  100\%       & 56,395 & 364,232 & 112,790 & 27,800 \\
\hline
\end{tabular}
\end{table}

\section{Evaluation}\label{evaluation}

To evaluate our models in the context of existing work, we compared the performance of DistilBERT, CodeBERT, and Gemma-2B against prior approaches including GPT-2, RoBERTa, MobileBERT, and LSTM across standard benchmarks like SARD and SeVC. Our experiments demonstrate that DistilBERT, when fine-tuned on vulnerability-aware program slices, achieved a peak F1 score of 98.8\% on a balanced full dataset (56K samples), surpassing previously reported performances for DistilBERT (88.86\% F1) and Accuracy of CodeBERT (89.23\%) on the SARD dataset. Notably, even under computational constraints using small, balanced subsets, DistilBERT and CodeBERT maintained strong F1 scores above 95\%, comparable to the best-performing models like TCN (93.59\% F1). However, when trained on small subsets and tested on the full dataset (~393K samples), performance dropped by 8–10\%, revealing generalization challenges similar to those faced by earlier methods. Our results validate the efficacy of lightweight transformer models and training strategies tailored for code semantics, offering competitive or superior performance at significantly reduced computational cost.

\begin{table}[h!]
\centering
\caption{Performance Comparison of VulDetect using GPT-2, CodeBERT, and LSTM on SARD and SeVC datasets}
\label{tab:vuldetect_results}
\begin{tabular}{|c|c|c|c|}
\hline
\textbf{Dataset} & \textbf{Model} & \textbf{Acc (\%)} & \textbf{F1 (\%)} \\
\hline
\multirow{3}{*}{SeVC~\cite{VulDetect} }
& GPT-2     & 87.63 & N/A  \\
& CodeBERT  & 83.23 & N/A  \\
& LSTM      & 71.59 & N/A  \\
\hline
\multirow{3}{*}{SARD~\cite{VulDetect}}
& GPT-2     & 92.59 & N/A  \\
& CodeBERT  & 89.28 & N/A   \\
& LSTM      & 76.86 & N/A  \\
\hline
\multirow{1}{*}{SARD~\cite{VulD-Transformer}}
& VulD-Transformer   & 87.5 & 87.5 \\
\hline
\multirow{3}{*}{SARD~\cite{BBVD}}
& RoBERTa     & 92.42 & 89.37   \\
& DistilBERT  & 95.39 & 88.86  \\
& MobileBERT      & 95.02 & 88.38   \\
\hline

\multirow{1}{*}{SARD~\cite{TCN}}
& TCN    & 93.54 & 93.59  \\

\hline
\end{tabular}
\end{table}

\subsection{Strategy 1: Full Vulnerability Training with Balanced Sampling: Train on all available vulnerable samples with an equal number of non-vulnerable samples; split 80\% for training and 20\% for testing.}

\begin{table}[htbp]
\caption{Model Performance Comparison on 56k Samples}
\centering
\scriptsize
\renewcommand{\arraystretch}{1.2}
\begin{tabular}{|l|c|c|c|c|c|c|}
\hline
\textbf{Category} & \textbf{Recall} & \textbf{Spec.} & \textbf{Prec.} & \textbf{F1} & \textbf{MCC} & \textbf{Acc.} \\
\hline
\multicolumn{7}{|c|}{\textbf{DistilBERT}} \\
\hline
PU & 99.14 & 98.62 & 98.60 & 98.87 & 97.75 & 98.87 \\
API & 99.60 & 98.03 & 98.08 & 98.83 & 97.65 & 98.81 \\
AU & 99.41 & 98.12 & 98.16 & 98.78 & 97.54 & 98.76 \\
AE & 98.61 & 98.29 & 98.31 & 98.46 & 96.90 & 98.45 \\
\textbf{Overall} & \textbf{99.27} & \textbf{98.36} & \textbf{98.37} & \textbf{98.82} & \textbf{97.64} & \textbf{98.81} \\
\hline
\multicolumn{7}{|c|}{\textbf{CodeBERT}} \\
\hline
PU & 97.36 & 96.49 & 96.46 & 96.91 & 93.85 & 96.92 \\
API & 98.39 & 91.15 & 91.85 & 95.00 & 89.81 & 94.79 \\
AU & 98.86 & 94.36 & 94.64 & 96.71 & 93.33 & 96.62 \\
AE & 95.84 & 89.88 & 90.54 & 93.11 & 85.89 & 92.87 \\
\textbf{Overall} & \textbf{97.81} & \textbf{94.43} & \textbf{94.60} & \textbf{96.18} & \textbf{92.30} & \textbf{96.12} \\
\hline
\multicolumn{7}{|c|}{\textbf{Gemma-2B}} \\
\hline
PU & 98.90 & 93.01 & 93.48 & 96.11 & 92.10 & 95.97 \\
API & 99.09 & 95.69 & 95.86 & 97.45 & 94.85 & 97.40 \\
AU & 97.23 & 92.83 & 93.21 & 95.17 & 90.17 & 95.04 \\
AE & 98.15 & 96.56 & 96.56 & 97.35 & 94.72 & 97.35 \\
\textbf{Overall} & \textbf{98.46} & \textbf{95.34} & \textbf{95.46} & \textbf{96.94} & \textbf{93.84} & \textbf{96.90} \\
\hline
\end{tabular}
\label{tab:comparison_singlecol}
\end{table}

\subsection{Strategy 2: Minimal Balanced Subset Training: Select the smallest number of vulnerable samples across all categories, match with equal non-vulnerable samples, and split 80\% for training and 20\% for testing.}

\begin{table}[htbp]
\caption{Model Performance (12k Samples) — Balance by Smallest Subclass}
\centering
\scriptsize
\renewcommand{\arraystretch}{1.2}
\begin{tabular}{|l|c|c|c|c|c|c|}
\hline
\textbf{Category} & \textbf{Recall} & \textbf{Spec.} & \textbf{Prec.} & \textbf{F1} & \textbf{MCC} & \textbf{Acc.} \\
\hline
\multicolumn{7}{|c|}{\textbf{DistilBERT}} \\
\hline
AE & 97.38 & 95.30 & 95.66 & 96.51 & 92.75 & 96.37 \\
PU & 94.14 & 94.83 & 94.70 & 94.42 & 88.99 & 94.49 \\
API & 96.48 & 91.28 & 91.70 & 94.03 & 87.87 & 93.88 \\
AU & 96.65 & 96.25 & 96.37 & 96.51 & 92.90 & 96.45 \\
\textbf{Overall} & \textbf{96.19} & \textbf{94.37} & \textbf{94.57} & \textbf{95.37} & \textbf{90.59} & \textbf{95.29} \\
\hline
\multicolumn{7}{|c|}{\textbf{CodeBERT}} \\
\hline
AE & 98.07 & 95.15 & 95.56 & 96.80 & 93.34 & 96.66 \\
PU & 94.14 & 95.12 & 94.98 & 94.56 & 89.28 & 94.64 \\
API & 95.92 & 91.70 & 92.03 & 93.93 & 87.69 & 93.81 \\
AU & 97.53 & 94.89 & 95.17 & 96.33 & 92.49 & 96.23 \\
\textbf{Overall} & \textbf{96.43} & \textbf{94.19} & \textbf{94.42} & \textbf{95.41} & \textbf{90.67} & \textbf{95.32} \\
\hline
\multicolumn{7}{|c|}{\textbf{Gemma-2B}} \\
\hline
AU & 97.52 & 94.86 & 95.28 & 96.39 & 92.47 & 96.23 \\
PU & 93.56 & 95.70 & 95.52 & 94.53 & 89.29 & 94.64 \\
API & 95.35 & 93.11 & 93.25 & 94.29 & 88.48 & 94.23 \\
AU & 97.09 & 96.40 & 96.53 & 96.81 & 93.50 & 96.75 \\
\textbf{Overall} & \textbf{95.90} & \textbf{94.99} & \textbf{95.12} & \textbf{95.51} & \textbf{90.90} & \textbf{95.45} \\
\hline
\end{tabular}
\label{tab:combined_14k_clean}
\end{table}

\subsection{Strategy 3: Full Test Evaluation: Generalization Evaluation on Full Test Set: Train on a minimal, balanced subset (as in Strategy B), but evaluate on the full remaining dataset to assess model generalization.}

\begin{table}[htbp]
\caption{Model Performance Comparison — Train 28k, Test 393k}
\centering
\scriptsize
\renewcommand{\arraystretch}{1.2}
\begin{tabular}{|l|c|c|c|c|c|c|}
\hline
\textbf{Category} & \textbf{Recall} & \textbf{Spec.} & \textbf{Prec.} & \textbf{F1} & \textbf{MCC} & \textbf{Acc.} \\
\hline
\multicolumn{7}{|c|}{\textbf{DistilBERT}} \\
\hline
API & 98.09 & 95.14 & 83.98 & 90.49 & 88.21 & 95.75 \\
AE & 99.48 & 95.74 & 81.84 & 89.80 & 88.22 & 96.34 \\
AU & 99.32 & 95.45 & 88.81 & 93.77 & 91.61 & 96.48 \\
PU & 95.83 & 97.97 & 84.07 & 89.57 & 88.55 & 97.75 \\
\textbf{Overall} & \textbf{97.22} & \textbf{97.30} & \textbf{84.86} & \textbf{90.62} & \textbf{89.33} & \textbf{97.29} \\
\hline
\multicolumn{7}{|c|}{\textbf{CodeBERT}} \\
\hline
API & 97.26 & 96.99 & 89.34 & 93.13 & 91.38 & 97.04 \\
AE & 99.37 & 97.50 & 88.49 & 93.62 & 92.52 & 97.81 \\
AU & 98.87 & 96.80 & 91.81 & 95.21 & 93.50 & 97.35 \\
PU & 94.42 & 98.75 & 89.40 & 91.84 & 90.94 & 98.31 \\
\textbf{Overall} & \textbf{96.20} & \textbf{98.31} & \textbf{89.83} & \textbf{92.91} & \textbf{91.83} & \textbf{98.02} \\
\hline
\multicolumn{7}{|c|}{\textbf{Gemma-2B}} \\
\hline
API & 98.10 & 95.78 & 85.77 & 91.52 & 89.47 & 96.25 \\
AE & 99.71 & 96.33 & 84.00 & 91.18 & 89.79 & 96.88 \\
AU & 99.08 & 95.45 & 88.77 & 93.64 & 91.42 & 96.42 \\
PU & 95.42 & 97.99 & 84.20 & 89.46 & 88.41 & 97.73 \\
\textbf{Overall} & \textbf{96.97} & \textbf{97.43} & \textbf{85.42} & \textbf{90.83} & \textbf{89.54} & \textbf{97.37} \\
\hline
\end{tabular}
\label{tab:train28k_test393k}
\end{table}

\begin{figure}[h]
    \centering
    \includegraphics[width=9cm, height=6cm]{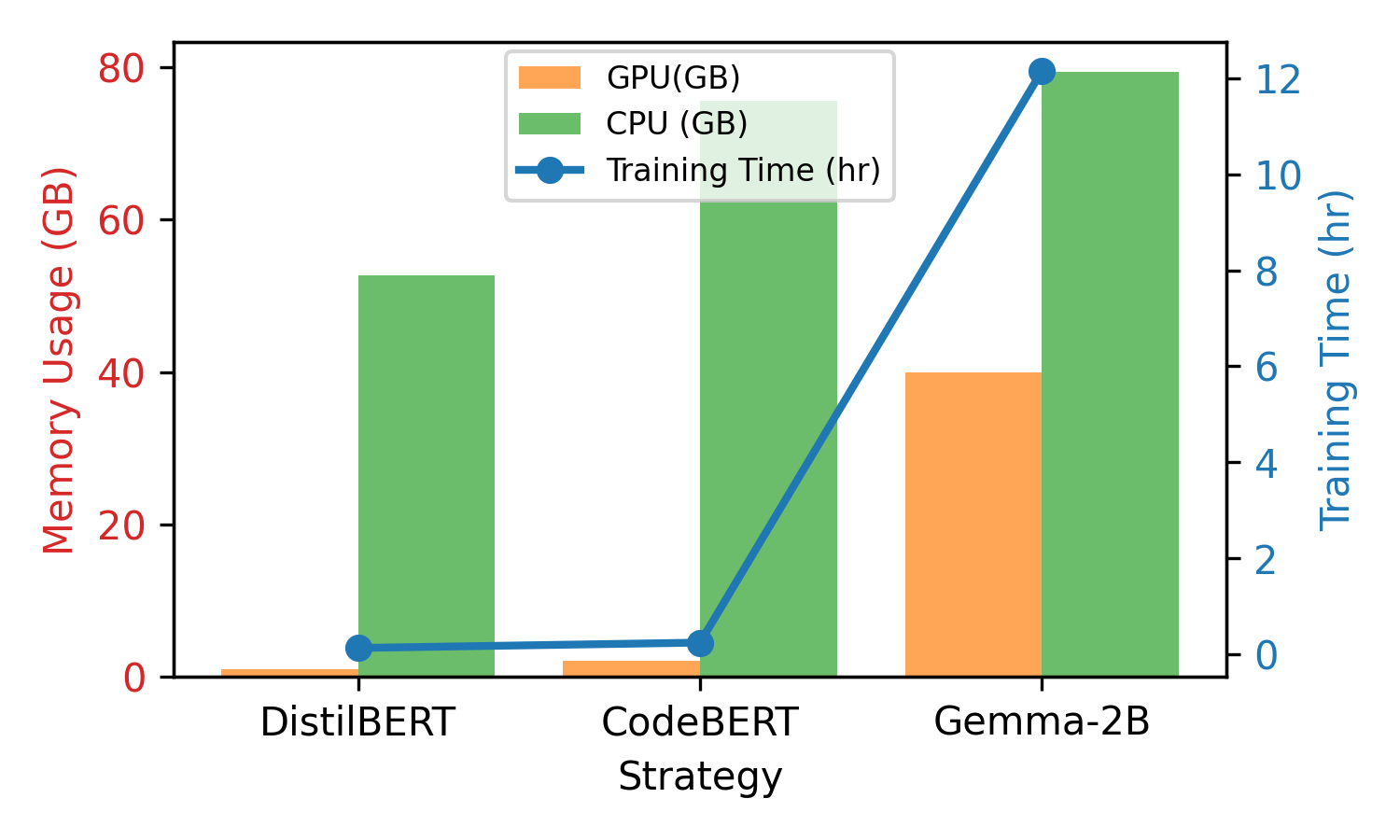}
    \caption{Trained on 28k and tested on remaining data}
    \label{fig:Trained on 28k and tested on remaining data}
\end{figure}

\begin{figure}[h]
    \centering
    \includegraphics[width=9cm, height=6cm]{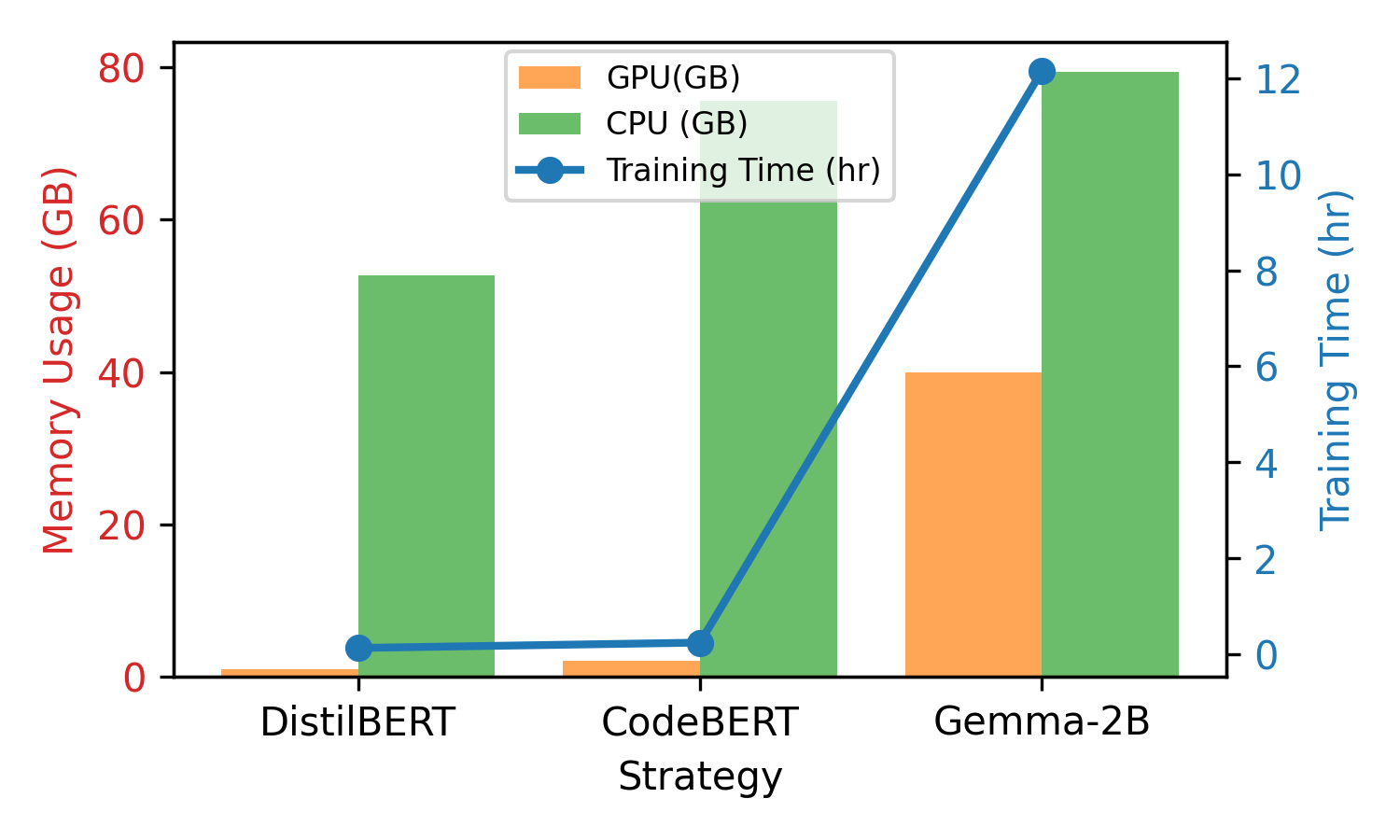}
    \caption{Trained on 28k and tested on remaining data}
    \label{fig:Trained on 22k and tested on 6k}
\end{figure}

\begin{figure}[h]
    \centering
    \includegraphics[width=9cm, height=6cm]{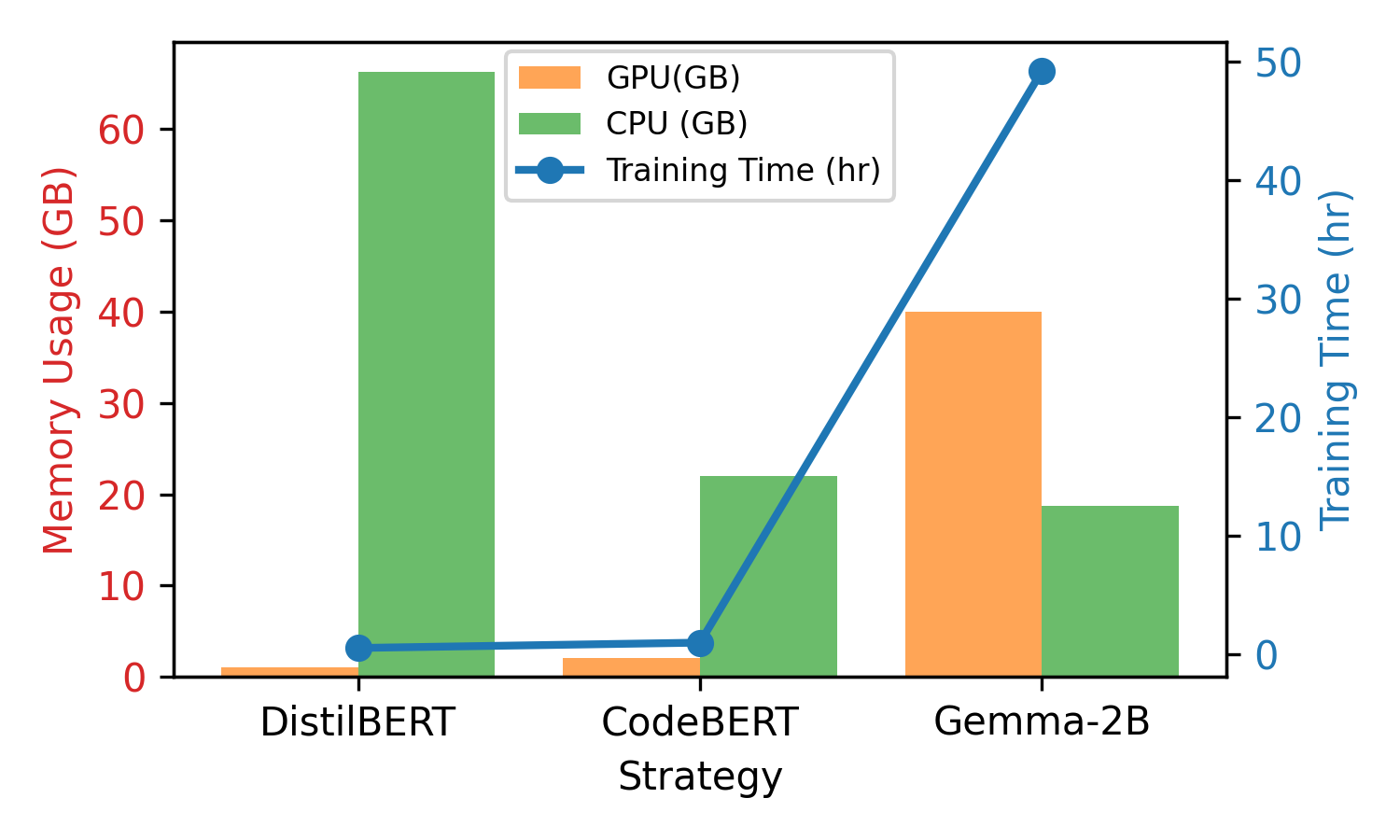}
    \caption{Trained on 44k and tested on 10k}
    \label{fig:Trained on 56k and tested on 20k}
\end{figure}

\section{Discussion}\label{discussion}

We compare three training strategies for vulnerability detection using three models of varying size: DistilBERT, CodeBERT, and Gemma-2B. Each strategy differs in the amount and sampling method of the training data, ranging from a minimal balanced subset to the full vulnerability dataset. We describe the dataset design, train/test split, compute resource usage (memory and training time), and F1-score results for each model under each strategy. The results demonstrate trade-offs between data quantity, model size, and computational cost, highlighting that even small models trained on limited data can achieve strong performance, while larger models and full datasets yield only moderate gains at significantly higher cost.

\subsection{Experimental Strategies}

\subsubsection{Strategy 1: Full Vulnerability Training with Balanced Sampling}
In Strategy 1, we addressed data imbalance by selecting a small, balanced subset from the dataset. Specifically, we took the minimum available vulnerable samples (3,475) and matched them with an equal number of non-vulnerable samples from each data kind (FC-kind, AU-kind, PU-kind, AE-kind). This dataset was split into 80\% for training and 20\% for testing. 
The computational cost was low: DistilBERT and CodeBERT completed training in approximately 0.13--0.24 hours, while the larger Gemma-2B model required around 12.1 hours. GPU memory usage was minimal for DistilBERT (1.05 GB) and CodeBERT (2.05 GB), but significantly higher for Gemma-2B (around 40 GB). CPU memory usage was moderate to high, with DistilBERT and CodeBERT using 52--75 GB and Gemma-2B around 79 GB. 
The models achieved strong F1 Scores: 95.3\% for DistilBERT, 95.4\% for CodeBERT, and 95.5\% for Gemma-2B. These results demonstrate that even a small, well-balanced dataset can lead to strong model performance with relatively low computational demands.

\subsubsection{Strategy 2: Minimal Balanced Subset Training}
In Strategy 2, we used all available vulnerability samples (56,395) and an equal number of non-vulnerable samples to create a much larger, balanced dataset. This dataset was then split into 80\% for training and 20\% for testing.
Due to the increased dataset size, training times rose significantly: DistilBERT and CodeBERT trained in approximately 0.54 and 0.98 hours respectively, while Gemma-2B took about 49.2 hours. GPU memory consumption remained similar, with smaller models using around 1--2 GB and Gemma-2B using approximately 40 GB. CPU memory requirements also stayed relatively moderate, around 18--66 GB across all models. 
Importantly, F1 Scores improved dramatically: DistilBERT achieved 98.8\%, CodeBERT reached 96.1\%, and Gemma-2B scored 96.9\%. These results clearly demonstrate that using a larger and richer training dataset leads to significant improvements in model performance, even though it increases computational cost.

\subsubsection{Strategy 3: Full Test Evaluation: Generalization
Evaluation on Full Test Se}
In Strategy 3, we maintained the same balanced training set as Strategy 1 (3,475 vulnerable and 3,475 non-vulnerable samples), but evaluated the models on the entire remaining dataset (approximately 364,000 samples). 
Training times were slightly longer than Strategy 1, with DistilBERT and CodeBERT training in 0.16--0.30 hours, and Gemma-2B requiring approximately 15.1 hours. GPU memory usage remained similar, while CPU memory requirements varied greatly: DistilBERT needed around 126 GB, whereas CodeBERT and Gemma-2B were more efficient with 72 GB and 13 GB respectively. 
Despite the manageable computational cost, F1 Scores dropped compared to Strategy 1: 90.6\% for DistilBERT, 92.9\% for CodeBERT, and 90.8\% for Gemma-2B. This indicates that while training on a small, balanced dataset is effective, models struggle to generalize when exposed to a large, complex, and diverse real-world dataset during testing.

\subsection{Comparative Analysis}
Across all three strategies, a clear trade-off is observed between dataset size, computational cost, and model performance. Strategy 1 provided strong results with very low training time and memory usage, but was limited by the small training set. Strategy 3 highlighted that training on small data does not generalize well to large real-world datasets, even though computational requirements were still manageable. Strategy 2, while requiring the highest computational resources (especially for the Gemma-2B model), achieved the best F1 Scores due to the larger and more diverse training set. Overall, these findings suggest that for software vulnerability detection, investing in larger, high-quality datasets is more beneficial than solely relying on larger models or computational optimizations.

\subsection{When to Use Which Strategy}
The choice of training strategy should balance accuracy needs and resource availability:
\begin{itemize}
    \item \textbf{Strategy 1} is ideal when compute resources are limited or quick iterations are needed.
    \item \textbf{Strategy 2} is appropriate for scenarios needing moderate improvements without overwhelming resource consumption.
    \item \textbf{Strategy 3} is suitable when maximum possible F1-score is required and sufficient infrastructure is available.
\end{itemize}

\subsection{Key Takeaways and Interpretations}
\subsubsection{Data set selection and balancing vs. model size and data set size}

Across these three strategies, we observe a clear tradeoff between model complexity and detection accuracy. Using a small, balanced dataset with DistilBERT (Strategy 1) is efficient in runtime overhead and resource usage. It drastically reduces training time and memory consumption while still achieving high F1 scores that are within reach of larger models. Strategy 1 demonstrates that \textit{careful data set selection and data set balancing can compensate for the limitations of smaller model sizes}, allowing a 66M-parameter model to perform nearly on par with models 2–30× larger.

Increasing the size of models and the size of training data (Strategy 2) will improve the detection accuracy, as CodeBERT’s higher capacity and pre-training in source code yield a notable boost in F1. However, the computational cost is disproportionately higher than the accuracy gain. Finally, scaling up to a 2B-parameter LLM. Strategy 3 yields only minor accuracy improvement beyond the CodeBERT-level baseline, yet incurs a disproportionately high computational expense.

\subsubsection{Hyperparameter fine-tuning vs. model size and data set size}

Our results suggest that \textit{scaling models yields modest gains compared to the benefits gained from balanced data and efficient fine-tuning}, pointing to the importance of smart dataset design and right-sized modeling in achieving both high accuracy and computational efficiency. Excessively large models may only be justified when the absolute highest performance is required, since they demand exponentially more compute. \textit{For vulnerability detection, an intermediate-sized model with a well-curated training set can achieve an excellent balance of accuracy and efficiency}.

\section{Conclusion}\label{conclusion}

Our findings demonstrate that lightweight transformer models like DistilBERT, when trained on carefully balanced datasets, achieve near-optimal vulnerability detection performance, a 98.8\% F1 score, with significant savings on computational cost -- 40× faster training time and 20× lower memory usage compared to larger models like Gemma-2B. While scaling model size and data set size yields marginal gains in vulnerability detection accuracy, +2\% F1 for CodeBERT, at disproportional computation cost, 49.2 GPU hours for CodeBERT vs. 0.54 GPU hours for DistilBERT. This indicates that smaller models are more efficient for practical deployment. Crucially, our results show that data balancing is as impactful as model architecture. A 95\%+ F1 score is achieved on a small balanced data set with minimal computation cost. Our results suggest that real-world vulnerability detection should prioritize data set quality and model efficiency over model scaling.

\bibliographystyle{plain} 
\bibliography{ref}

\end{document}